# Two Photon excitation microscopy of individual Single-Walled Carbon Nanotubes

Nasim Akhtar[1,2], Marc Tondusson[1,2], Benjamin Flavel [3],
Stéphane Bancelin[1,2,*], and Laurent Cognet[1,2,*]

[1]*Laboratoire Photonique Numérique et Nanosciences, Université de Bordeaux, 33400 Talence, France,*

[2]*LP2N, Institut d'Optique Graduate School, CNRS UMR 5298, 33400 Talence, France*

[3]*Institute of Nanotechnology, Karlsruhe Institute of Technology, D-76131 Karlsruhe, Germany*

** correspondence: stephane.bancelin@u-bordeaux.fr, Laurent.cognet@u-bordeaux.fr*

**Abstract**

Two-photon fluorescence imaging achieves deep-tissue penetration through long excitation wavelengths and nonlinear excitation confinement. The NIR IIb transparency window is particularly attractive, as it optimally balances tissue scattering and absorption. However, efficient fluorophores for two-photon excitation in this window remain limited. Moreover, the weak near-infrared emission of individual emitters and the low photon detection efficiency have so far precluded single-particle imaging. Here, we characterize the two-photon excitation properties of chirality-sorted pristine and quantum color center-functionalized single-walled carbon nanotubes under 1630 -1700 nm excitation. By measuring and comparing their two-photon action cross-sections, we identify quantum color center-functionalized (6,5) nanotubes emitting at 1140 nm, as the most promising emitter, with an exceptionally large cross-section of (17.1 ± 0.6) x $10^3$ GM/nm. Leveraging these favorable photophysical properties, we image individual nanotubes in the NIR IIb window, to our knowledge the first demonstration of single-particle imaging in this spectral window. These results establish quantum color center-functionalized (6,5) nanotube as a strong candidate for long-wavelength two-photon imaging and pave the way toward single-particle imaging in scattering tissues.

Optical imaging is essential in biological research and diagnostics, enabling the visualization of sub-cellular morphology and functional dynamics of biological structures within complex biological systems in a minimally invasive approach. However, recording high-resolution images in thick tissues is challenging because light scattering and absorption critically limit imaging depth. Multiphoton imaging techniques such as two-photon (2P) and three-photon (3P) excited fluorescence address these limitations by shifting fluorescence excitation to longer wavelengths, where reduced scattering and absorption enhance light penetration [1]. Over the past decades, multiphoton microscopy has advanced considerably and now stands as the gold standard for deep tissue imaging in optical microscopy [2,3]. Throughout this work, we follow the near-infrared window nomenclature commonly used in bioimaging: NIR-I (~700–1000 nm), NIR-II (1000–1700 nm) and its long-wavelength NIR-IIb sub-window (1500–1700 nm).

Two-photon excitation microscopy (2PEM) has traditionally exploited the biological transparency window (~800-1000 nm), typically using excitation wavelength around 920 nm for green fluorophores such as GFP and GCaMP [4]. This has enabled, for instance, *in vivo* imaging of neuronal activity in superficial cortical layers. Representative applications include calcium imaging in the mouse visual cortex and dendritic spine imaging [5]. Subsequently, excitation at longer wavelengths in the ~1300 nm range, which corresponds to the NIR-II window, was introduced for red fluorophores (e.g. tdTomato, mCherry, jRGECO1a), allowing deeper imaging in 2PEM owing to reduced light scattering. These approaches have been used for dual-color imaging of neural circuits and deeper cortical imaging [6,7]. To achieve even greater imaging depths, three-photon excitation microscopy (3PEM) was subsequently introduced, exploiting excitation ≃1300 nm, for imaging green fluorophores such as GCaMP. This enabled functional imaging in deep cortical layers and in the hippocampus beyond the depth limit of conventional 2PEM, often reaching 1 mm or more below the brain surface [8].

Lastly, balancing tissue scattering and absorption, the optimal wavelength for deep penetration in biological tissues has been identified in the NIR IIb region, around 1700 nm [9]. At this wavelength, 3PEM enables the use of common red fluorescent dyes and fluorescent proteins, such as mCherry and tdTomato, as well as red-shifted calcium indicators like jRGECO1a [3,10].

In contrast, the use of the NIR IIb transparency window for 2PEM remains limited, with only a handful of studies demonstrating its feasibility, notably using indocyanine green (ICG) [11], which can be attributed to the lack of efficient fluorophores in the NIR-II spectral window. This observation calls for further development of novel fluorescent molecules specifically suited for efficient 2P excitation at longer wavelengths.

In parallel, single-walled carbon nanotubes (SWCNTs) have emerged as highly promising near-infrared (NIR) fluorophores. Notably, their emission bands lie within the NIR-II region [12]. In addition to these favorable spectral properties, SWCNTs exhibit exceptional photostability compared to conventional fluorophores, enabling prolonged imaging with minimum photobleaching and photodamage[13-15]. The optical properties of SWCNTs are governed by their chirality (n,m), with each semiconducting species exhibiting distinct optical transitions. While all semiconducting SWCNTs emit within the NIR region, each chirality is associated with well-defined excitation and emission wavelengths [16]. Recent studies have demonstrated that the brightness of SWCNTs can be markedly enhanced through chemical functionalization, introducing local color centers (CCs), also referred to as quantum defects [17-19]. SWCNTs incorporating color centers (CCNTs), which exhibit strong absorption and emission in the NIR-II spectral window [20], are emerging as powerful tools for biological imaging and sensing in complex biological environments.

2P excitation spectroscopy has been instrumental in establishing the excitonic origin of optical transitions in SWCNTs, reflecting the presence of a rich excitonic manifold that allows 2P transitions associated with the so-called $E_{11}$ transition [21,22]. Interestingly, for specific chiralities, these 2P allowed transitions correspond to excitation wavelengths in the NIR IIb window, thereby fitting well within the optimal transparency imaging window. Indeed, a recent study reported the use of SWCNTs as fluorescent emitters for 2PEM with a 1640 nm excitation wavelength [23].

While 2P imaging of single quantum dots has been demonstrated owing to their large 2P action cross-sections [24], multiphoton fluorescence imaging at the single-particle level remains extremely challenging due to the intrinsically low 2P absorption action cross-sections and limited photostability of most fluorophores [25–28]. Here, we hypothesize that single particle detection using 2PEM with SWCNTs may be achievable, owing to their high intrinsic brightness, which can be further enhanced through quantum defect engineering in CCNTs [17-19]. These developments make carbon nanotubes particularly attractive to overcome a key limitation in single-particle multiphoton imaging: the inevitable trade-off between the high excitation intensities required for nonlinear processes and the bright, photostable emission needed for reliable single-particle detection.

## 2P Excitation Spectroscopy of SWCNTs in the NIR IIb window

In SWCNTs, only singlet excitons are optically active due to weak spin–orbit coupling, such that their transitions are governed by electric-dipole selection rules. For transitions polarized along the nanotube axis, one-photon (1P) absorption requires the initial and final states to have opposite symmetry, such that odd-parity (u-symmetry) excitons, including the lowest bright state (1u), are involved in such processes. In contrast, 2P excitation is only allowed if the final state shares the same parity as the initial state leading thus to the excitation of 2g excitons [29]. As a result, 1P transitions involve the lowest-lying 1u exciton, while 2P transitions exclusively access higher-energy exciton states (Figure 1a). Following 2P absorption, rapid relaxation to the lowest 1u exciton state occurs, resulting in NIR photoluminescence [21,22].

Since 2P excitation follows specific selection rules, the optimal chirality to be used as a fluorescent probe upon 2P excitation in the NIR IIb window cannot be directly inferred from 1P excitation properties, but instead, must be experimentally identified. To this end, a multi-chirality SWCNT sample (generated by the HiPco method) was prepared by dispersing raw material in 1% sodium deoxycholate (DOC) surfactant solution. This surfactant is known to provide bright and stable suspensions with a broad distribution of nanotube chiralities [30]. The sample composition was first characterized by acquiring 2D photoluminescence (excitation–emission) map using 1P excitation (Figure 1b).

We then employed a tunable femtosecond laser source (emission wavelength between 1630 ± 7 nm and 1700 ± 7 nm, 4 MHz repetition rate, ≃ 80 fs pulse duration), a cost-effective pulsed laser specifically designed and produced for this study by ALPHANOV (Talence, France), to systematically evaluate chirality-dependent 2P excitation at 1660 nm and to identify a suitable candidate for 2PEM (Figure 1c). The excitation efficiency was further quantified by measuring the fluorescence intensity as a function of excitation wavelength under constant excitation power (20 mW before the focusing lens, figure 1d)

We observed that (7,5) nanotubes reach a peak excitation at 1670 ± 5 nm, consistent with previously reported excitation wavelengths for this chirality, well within our laser range. In contrast, although (6,5) and (7,6) nanotubes are efficiently excited, their peak excitation wavelengths were out of our range, below 1640 nm and above 1700 nm, respectively [22]. However, quantitative comparison of the fluorescence intensity between these 3 chiralities further requires normalization to their respective concentrations. Due to the presence of many chiralities in HiPco sample with absorption peaks in the vicinity, it turned out to be difficult to accurately determine their precise concentration. We thus employed highly pure single-chirality solutions (both (6,5) and (7,5)) [31] and determined their concentrations using UV–vis absorption spectroscopy before comparing their emission profiles upon 2P excitation. Figure 1e shows the fluorescence intensities normalized to the nanotube concentrations. It appears that (6,5) nanotubes generate twice as much fluorescence as (7,5) nanotubes, as measured from the curve integral, even when not excited at resonance. This indicates that the 2P action cross-section of (6,5) nanotubes is at least twice as large as that of (7,5) nanotubes, in accordance with 1P measurements [32]. Power-dependent fluorescence measurements performed at 1670 nm for (7,5) nanotubes and at 1640 nm for (6,5) nanotubes revealed a near-quadratic dependence, with slope of 1.88 ± 0.07 and 1.85 ± 0.09 respectively, confirming that the signal originates from a second-order process (Figure 1f). It should be noted that at high excitation power (> 20 mW), the 2P fluorescence intensity starts to saturate and progressively deviates toward a linear dependence, as also observed upon 1P excitation [33]. This behavior can be attributed to a significant increase in the exciton population, leading to ultrafast exciton–exciton annihilation [34,35]. Consequently, newly generated excitons are non-radiatively quenched leading to a reduction in fluorescence efficiency.

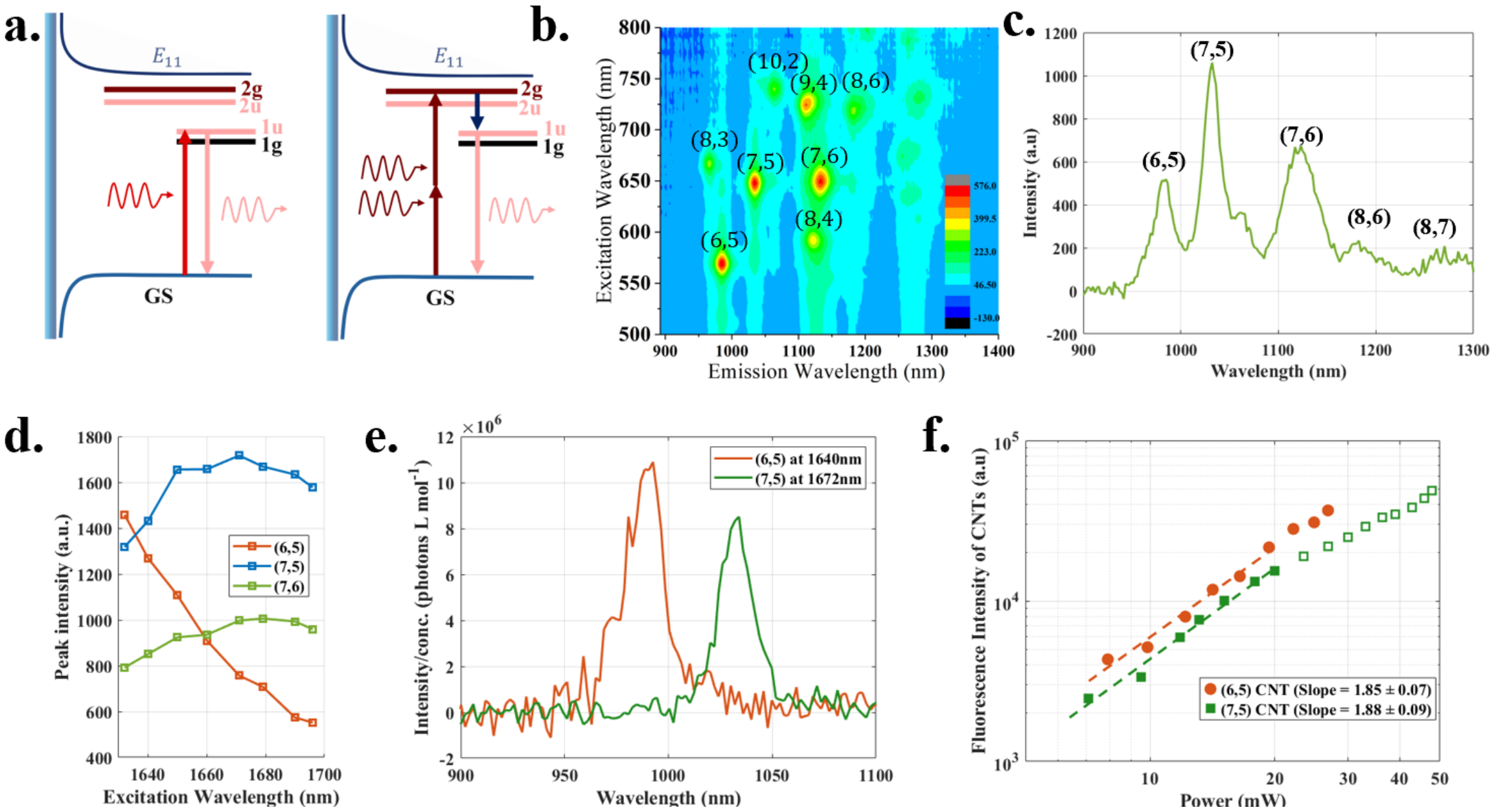

*Figure 1: a) in SWCNTs, 1P excitation couples to odd-parity (u) excitons, while even-parity (g) states are optically forbidden (left panel). In contrast, 2P excitation enables access to even-parity (g) exciton states prior to relaxation and near-infrared emission (right panel). b) 1P excitation 2D photoluminescence map (excitation–emission profile) of SWCNTs (HiPco) with corresponding nanotubes chiralities. c) 2P emission spectra of the multi-chirality HiPco SWCNTs in 1% DOC, using 1660 nm excitation. d) Fluorescence intensity of (6,5) SWCNTs, (7,5) SWCNTs, and (7,6) SWCNTs for different excitation wavelengths with fixed 20 mW excitation power. e) 2P fluorescence intensity normalized to their respective concentration at fixed 20 mW excitation power for chirality sorted (7,5) SWCNT at 1670 nm and (6,5) SWCNTs at 1640 nm. f) Log–log plot of 2P excited fluorescence as a function of excitation power for chirality sorted (7,5) SWCNT at 1670 nm and (6,5) SWCNTs at 1640 nm. Dotted lines are linear fits to the experimental results (markers). The slopes are indicated in the legend, revealing a near-quadratic dependence of the emission intensity on excitation power for both chiralities.*

## 2P Microscopy of multi-chirality SWCNT samples

We next evaluated the feasibility of imaging individual SWCNTs by 2PEM. To this end we used a custom-built multiphoton microscope (Figure 2a) based on the femtosecond laser mentioned above and with optimized PMT for NIR detection. We first performed imaging experiments using multi-chirality HiPco SWCNTs serving as a baseline sample to probe the intrinsic photophysical behavior of pristine nanotubes under 2P excitation. SWCNTs spin-coated onto poly-L-lysine-coated glass coverslips, as described in the Methods, were first imaged by one-photon excitation microscopy (1PEM) using continuous-wave 845 nm excitation from a Ti:Sapphire laser with 8 kW/cm$^2$ and 1 ms dwell time, revealing individual nanotubes well separated on our sample (Figure 2b). We then imaged the same field of view with 2PEM at low power (0.3 mW after the objective lens, 1640 nm) and long dwell times (20 ms) revealing weak, yet detectable, fluorescence signals that can be assigned as originating from isolated individual nanotubes. To support this assignment, we added an IR spectrometer in the detection path, and we recorded 1P emission spectra at different spot, which exhibit the characteristic features of individual nanotubes (Figure 2c) [36]. The observation of multiple, well-defined monomodal emission spectra from a multi-chirality sample indeed confirms that individual nanotubes are being detected. Notably, the 1P fluorescence image shows distinct spatial localization of specific SWCNT chiralities: (6,5) in the blue circle, (10,2) in the orange circle, and a mixture of (7,6) and (8,4) in the green circle.

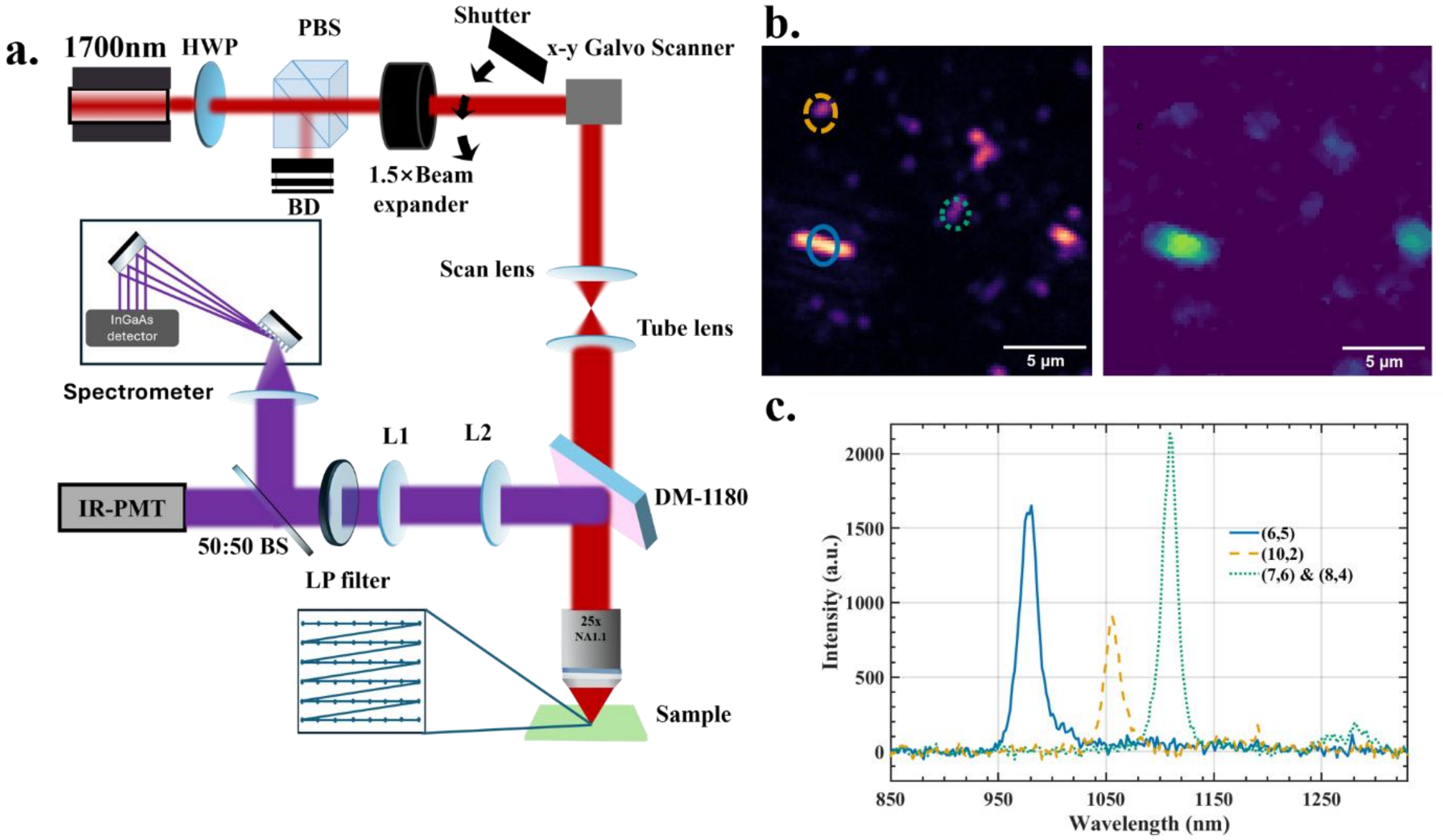

*Figure 2: a) Schematic of the multiphoton imaging setup. PBS: polarization beam splitter, BD: beam dump, DM: dichroic mirror, IR-PMT: infrared photomultiplier tube, HWP: half-wave plate, BE: beam expander, LP Filter: 950 nm long-pass filter. b) 1P (left) and 2P (right) imaging of SWCNTs deposited on a coverslip. 1P images were acquired using 845 nm excitation while 2P images were obtained with 1640 nm excitation. c) 1P emission spectrum acquired from the individual SWCNTs labelled in b).*

Since a nanotube longer than the point spread function behaves as a line object, the intensity profile taken perpendicular to its axis directly gives an estimate of the lateral resolution. Gaussian fits of transverse profiles across 20 nanotubes yield FWHM = 800 ± 100 nm under 1P excitation at 845 nm and 1100 ± 100 nm under 2P excitation at 1640 nm. For comparison, the diffraction limited FWHM of the excitation PSF with our 1.1 NA objective is 390nm for 1P at 845nm and 540nm for 2P at 1640nm. Both measured widths are thus about 60% larger than the theoretical values, but their ratio (1.4 ± 0.2) agrees with the theoretical ratio (1.37) at equal NA. The lower resolution of 2P images is therefore fully accounted by the longer excitation wavelength. Noteworthy, the larger FWHM, observed for both modalities, can be attributed to lower effective numerical aperture due to underfilling of the back-aperture of the objective. This has been chosen to compensate the limited available power with our laser source by maximizing the overall transmission. Future optimization could be done by using the full NA through overfilling the back aperture of the objective and correcting for residual aberrations. Further improvement could be done by exploiting the shorter emission wavelength through confocal and image-scanning detection or super-resolution approaches, at the cost of lowering photon budget.

While these observations demonstrate 2PEM of single SWCNTs, we noticed that only rather long nanotubes could be identified, while abundant shorter tubes, observed in 1PEM, were not bright enough for single-nanotube 2PEM detection.

## 2PEM of pure (6,5) CCNTs

To detect small individual SWCNTs, stronger fluorescence emission or higher excitation efficiency is required, as weakly emitting chiralities only provide barely detectable signals at the single-nanotube level. Recent studies have shown that SWCNT brightness can be significantly increased through chemical functionalization to implant local color centers (CCs), also called quantum defects [17-19]. Among them, oxygen functionalization can generate such CCs by introducing $sp^3$ defects that locally modify the nanotube electronic structure and create a lower-energy, localized excitonic state $E_{11}^{*}$[18], as illustrated in Figure 3a. When mobile excitons diffuse into these CCs, they are locally trapped and recombine radiatively, resulting in red-shifted emission and high fluorescence intensity compared to pristine nanotubes [37]. Leveraging this key advantage, oxygen functionalized nanotubes (CCNTs) have been notably shown to be instrumental in mapping the brain extracellular space in living tissue using 1P microscopy [38,39].

In this context, we investigated the possibility of detecting individual (6,5) CCNTs using 2PEM. We first characterized a sample solution using our spectroscopy setup (see Methods). The observation of the characteristic red-shifted emission peak (at 1140 nm) reflects the presence of oxygen related CCs on the nanotubes (Figure 3b). As anticipated, we also observed that (6,5) CCNT solutions are significantly brighter under 2P excitation by at least a factor two, compared with pristine (6,5) SWCNTs, as determined by integrating the total number of photons under the spectral emission curve.

2P absorption in pristine SWCNTs is mainly dominated by the intrinsic $sp^2$ hybridized carbon lattice of the nanotube. At the low CC densities used to maintain high quantum efficiency of CCNTs, the presence of CCs does not significantly alter the nanotube absorption spectrum [40]. Indeed, CCs introduce localized exciton states that increase the fluorescence quantum yield by trapping excitons and reducing non-radiative recombination at quenching sites [17]. Consequently, although pristine SWCNT ensembles typically exhibit very low 1P quantum yields (~1.4%) [41], single (6,5) CCNTs can reach quantum yields of up to 20% [38]. We believe that exciton trapping at CCs is also at the origin of the observed brightness increase for CCNTs under 2PEM.

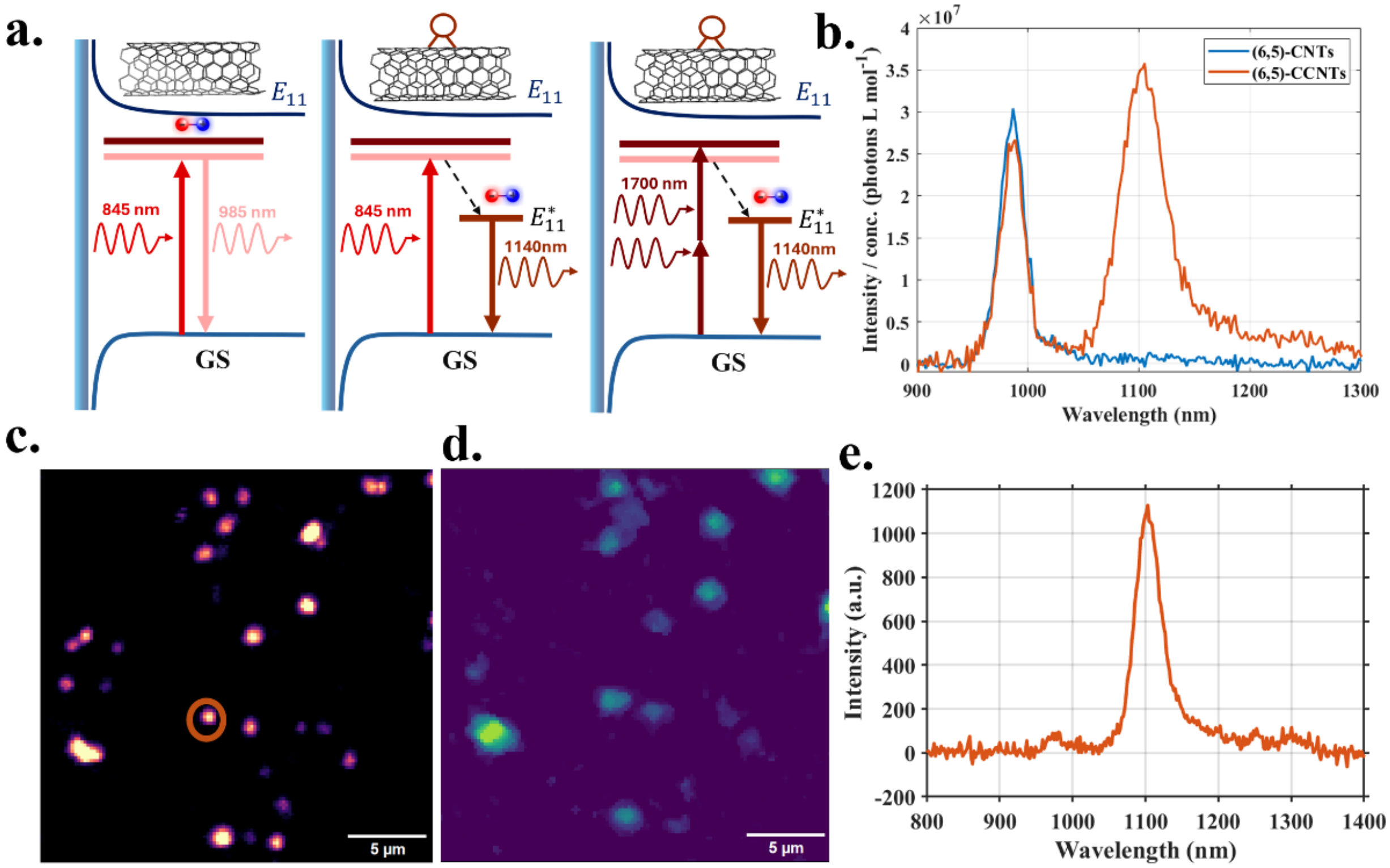

*Figure 3: a) Structures of pristine (left) and oxygen-functionalized (middle and right) SWCNTs and corresponding energy levels. Oxygen ($sp^3$) functional groups acting as organic color centers (ox-CCNTs) create*

*localized defect states that trap excitons and produce red-shifted emission. The left and middle panels show 1P excitation, whereas the right panel illustrates 2P excitation in the NIR IIb. b) 2P fluorescence spectra of pristine (6,5)-CNTs (blue trace) and oxidized (6,5)-ox-CCNTs (orange trace) under excitation at 1640 nm. The pristine (6,5)-CNTs show the $E_{11}$ emission peaks centered near 990 nm, whereas the oxidized sample exhibits an additional defect-state $E_{11}^{*}$ emission peak centered near 1140 nm. The y-axis represents the emission intensity normalized by concentration (photons.L.mol$^{-1}$) c) 1P imaging of (6,5) CCNTs deposited on a coverslip acquired using 845 nm excitation light. d) 2PEM images of the same field of view obtained with 1640 nm excitation. e) 1P emission spectrum acquired from an individual carbon nanotube labelled in c) using the microscopy setup.*

We finally investigated the possibility to image single CCNTs using our microscope, following the same approach as for pristine SWCNTs. We first acquired 1P images of sparsely dispersed individual (6,5) CCNTs, on a coverslip using 845 nm excitation, followed by 2PEM imaging of the same field of view (Figure 3 c and d). In these images, the CCNTs appeared as small, diffraction-limited fluorescence spots, each representing an individual nanotube, consistent with the resolution expected for a 2PEM system with 1640 nm excitation wavelength.

Lastly, to verify that the signal from each bright spot originated from individual (6,5) CCNTs and to distinguish them from possible bundles or different defect configurations, we recorded the emission spectrum of the same nanotube (Figure 3e). The measured spectrum matches the characteristic emission of (6,5) CCNTs, confirming that single CCNTs could be efficiently imaged by 2PEM. Interestingly, comparing CCNT images obtained from the two excitation modalities indicates that close to all single CCNTs detected by 1PEM are also present in 2PEM images.

## Measurement of 2P action cross-sections of SWCNTs

Using our laser-scanning microscope, the measured 2P action cross sections of all investigated samples, including the (6,5) CCNTs, were finally determined under identical experimental conditions, at low excitation power (1 mW at the sample) to remain in the unsaturated regime. As in Figure 1f, the epi-detected fluorescence intensity scaled quadratically with excitation power ($I \propto P^2$, data not shown), confirming that the signal originates from a 2P process. To extract the action cross-section ($\eta_2\sigma_2$), the detected fluorescence ($F_2$) was related to excitation and detection parameters through the following equation [27]:

$$\eta_2\sigma_2 = \frac{f\tau\lambda\pi F_2}{4g_p^2 \Phi C n P^2 T T_{cal}^2} \qquad (1)$$

where f is the laser repetition rate, τ the pulse duration, λ the excitation wavelength, P the excitation power before the objective lens, $T_{cal}$ the transmission factor converting this power into the power at focus (0.83), C the nanotube concentration, *n* the refractive index of the medium, Φ the collection efficiency of the path up to the PMT, $g_p^2$ the degree of temporal coherence (we used 0.59, the value for *sech*$^2$-shaped pulse [27]), and T is the integration time given by the pixel dwell time (30 ms). The concentration of each sample was determined independently from UV–Vis–NIR absorption spectrum using the absorption cross-sections per carbon atom reported in [32], and converted into a nanotube length concentration using the number of carbon atoms per unit length [42]. The action cross-sections are thus given per unit length of nanotube in GM/nm. The collection efficiency of the optical path preceding the PMT was calculated by weighting the fluorescence emission spectrum by the wavelength dependent transmission of every element in this path (objective, dichroic mirror, filter and relay-lenses) as well as the detection efficiency of the PMT, yielding $\Phi \simeq 15\%$ over the SWCNT emission range (14.9% at 1140 nm to 15.4% at 980nm).

The 2P action cross-sections obtained for all samples following this procedure are summarized in Table 1. Notably, the (6,5) CCNTs exhibit a pronounced enhancement in the 2P action cross-section compared to pristine (6,5) SWCNTs, exceeding them by more than a factor of three. This enhancement is consistent with the increased brightness observed in both spectroscopic and imaging measurements and is attributed to the localized nature of excitons at oxygen-related CCs, which induce redshifted emission and improve radiative recombination efficiency by suppressing the quenching pathways of mobile $E_{11}$ excitons.

*Table 1: Two-photon action cross sections of SWCNTs. Uncertainties correspond to the standard deviation over 12 repeated measurements.*

| Chirality | Excitation Wavelength (nm) | 2P Action Cross-Section ($\times 10^3$ GM nm$^{-1}$) |
|---|---|---|
| (6,5) CNTs | 1640 | 6.8± 2.1 |
| (7,5) CNTs | 1640 | 2.7 ± 0.2 |
| (7,5) CNTs | 1660 | 3.4± 0.6 |
| (6,5) CCNTs | 1640 | 17.1 ± 0.6 |

These results reveal a clear dependence of the 2P action cross-section on both chirality and excitation wavelength: among the pristine tubes, (6,5) SWCNTs display a significantly larger cross-section than (7,5) SWCNTs, in agreement with our spectroscopic data, their stronger emission under 2P excitation, and our 1P measurements.

These large 2P action cross-section stem from the nature of SWCNTs as extended 1D absorbers, with about 90 carbon atoms per nanometer for (6,5) nanotubes. Their optical response arises from tightly bound excitons. Similarly to 1P absorption, [32], the 2P absorption cross-section of an individual nanotube scales with its length and reaches very large values for nanotubes of a few hundred nanometers. Additionally, in CCNTs, exciton trapping at color centers further increases the fluorescence quantum yield, and hence the action cross-section [17,40].

It should be noted that the value we obtain for (6,5) SWCNTs exceeds previously reported values [43], which we attribute to differences in sample preparation and experimental conditions. In particular, earlier studies commonly employed nanotubes dispersed in SDS surfactant and excited at 1550 nm, whereas we use highly purified samples dispersed in DOC and excite at 1640 nm. Differences in nanotube synthesis and surfactant environment can strongly influence nanotube dispersion, exciton dynamics, and non-radiative decay pathways [44,45], while detuning the excitation from the 2P resonance further modulates the measured 2P action cross sections.

In this study, we systematically investigated the 2P excitation of SWCNTs in the NIR IIb spectral window. Through 2P spectroscopy experiments on multi-chirality SWCNTs samples, we compared the fluorescence intensities of different chiralities to identify those most suitable for imaging applications. We then performed 1PEM and 2PEM on multi-chirality SWCNTs, and found that only longest, and therefore brightest, individual nanotubes could be detected by 2PEM. Shorter nanotubes, although present, were not sufficiently bright for single-nanotube detection under 2P excitation. To overcome this challenge, we finally took advantage of recently introduced (6,5) CCNTs to allow efficient single-nanotube detection, demonstrating for the first time 2P imaging of individual carbon nanotubes in the NIR IIb excitation window.

Our study highlights the potential of CCNTs as bright fluorophores that can be excited in the NIR IIb transparency window and emit around 1140 nm, thereby combining the reduced scattering and low autofluorescence of NIR-IIb excitation with the efficient collection of NIR-II emission, which may be advantageous for future deep-tissue imaging applications. Their high 2P brightness further facilitates the detection of individual objects, making them promising candidates for future applications in biological imaging and single-particle tracking.

# Material and Methods

## Sample preparations

Chirality-sorted SWCNTs were obtained from HiPco by gel chromatography. The separation relies on differences in the interaction of surfactant-coated nanotubes with a Sephacryl gel matrix, enabling the selective enrichment of specific nanotube chiralities as described in [31]. HiPco SWCNTs (Batch 195.7, Rice University), as well as chirality-sorted (6,5), (7,5), and (6,5)-CCNTs, were dispersed in aqueous solution using 1% DOC surfactant [30]. For spectroscopic measurements, 100 $\mu$L of the nanotube suspension was placed into a quartz cuvette (Hellma Analytics). For microscopy measurements, glass coverslips were coated with poly-L-lysine, before depositing of the nanotubes and spin coating to obtain isolated nanotubes suitable for single-particle imaging, following established protocols [13, 33, 36, 38].

## Photoluminescence map

PL maps for HiPco nanotubes (1% DOC surfactant) were recorded by a NanoLog spectrofluorometer (HORIBA Jobin Yvon). The samples were excited with a 450 W xenon source dispersed by a double-grating monochromator (excitation and emission beam slit width of 10 nm). The PL spectra were recorded using a liquid – $N_2$ – cooled InGaAs line camera (Symphony II) with 10 s integration time. The emitted light passed through an 850 nm long-pass filter and a Czerny–Turner spectrograph before reaching the detector.

## UV–Vis–NIR absorption spectroscopy

UV–Vis–NIR absorption spectra in the wavelength range of 400–1200 nm were recorded for the chirality sorted (7,5) and (6,5) samples and for (6,5) CCNTs using a spectrophotometer (Cary 5000, Agilent Technologies). The measurements were carried out using a quartz cuvette with an optical path length of 3 mm.

## Excitation source

The 2P excitation experiments were performed by a femtosecond laser source prototype, developed by ALPhANOV (Talence, France), based on soliton self-frequency shift. In short, a femtosecond pulse laser beam at 1550 nm is used to pump a photonic fiber in which a part of the beam is red-shifted up to 1700 nm with $\simeq$ 80 fs pulse width and 4 MHz repetition rate. This configuration enabled tuning the excitation wavelength as a function of the power injected in the fiber. After collimation, the beam passes through a 1550 nm long-pass filter (BLP01-1550R, Semrock) to reject the remaining pump at 1550 nm. A combination of a half-wave plate and a polarization beam splitter (PBS) is used to tune the incident power on the sample. The beam waist (≃ 3 mm) was measured before the sample cuvette using the knife-edge method. The pulse duration was measured directly after the microscope objective. Signal were recollimated through a lens with 50 mm focal length and send to a commercial FROG.

## 2P spectroscopy setup

For spectroscopy measurements, the pulsed laser was focused into the sample using a 30 mm focal length planoconvex lens. The average power at the sample varies from 5 to 25 mW and spectra were acquired with 60 s integration time. Fluorescence emission photons from the sample were collected with a pair of 50 mm focal length planoconvex lenses and sent to a liquid-N2-cooled SpectraPro 2150i equipped with an InGaAs array detector (Princeton Instruments). The background signal was recorded under identical conditions and subtracted from all measurements.

## Microscopy imaging

2P imaging was performed using a custom-made multiphoton microscope. Excitation was provided by the ALPhANOV's femtosecond laser, as described above, and incident power was tuned using a half-wave plate and a PBS. After passing through a 1.5× beam expander and a mechanical shutter (SH05R/M, Thorlabs) the excitation beam was raster-scanned by two galvanometer-driven mirrors (Saturn 9, ScannerMAX). The scan mirrors were imaged onto the back aperture of the microscope objective with a magnification of 4 times. The beam then passed through a dichroic mirror (DMLP1180R, Thorlabs) and focused into the sample using high numerical aperture water immersion lens (Nikon, 25x/1.1 NA).

The fluorescence from the sample was epi-collected, relayed using a 300 mm lens and 50 mm planoconvex lens, filtered using a 950 nm long-pass filter (Thorlabs, FELH950) and detected on a PMT (H15620-45, Hamamatsu Photonics) well-adapted for NIR detection. Accounting for PMT, filter, and dichroic, the detected spectral range was 950-1200 nm. For complementary 1P microscopy, similar setup was used with an 845 nm continuous-wave laser derived from a Ti:Sapphire laser (Mira 900, Coherent) and a suitable dichroic mirror (FF930-SDi01, Semrock).

## Acknowledgments

This work received financial support from the European Research Council (ERC) Synergy Grant (951294), the Agence Nationale de la Recherche (ANR-24-CE09-2351-01), EUR Light&T (PIA3 Program, ANR-17-EURE-0027), the France-BioImaging National Infrastructure (ANR-10-INBS-04-01), and the IdEx Bordeaux (Grand Research Program GPR LIGHT).